\documentclass[12pt,dvipsnames]{article}

\usepackage[utf8]{inputenc}
\usepackage{amsmath}

\usepackage[margin=1in]{geometry}
\usepackage{titlesec}
\titleformat{\section}{\large \bfseries}{\thesection}{1em}{}

\usepackage{amssymb}

\usepackage{graphicx}

\usepackage{tikz}
\usetikzlibrary{arrows.meta}
\usetikzlibrary{positioning}

\usepackage{csquotes}
\usetikzlibrary{matrix}

\usepackage{tabularx}
\usepackage{array}
\usepackage{booktabs}
\usepackage{enumitem}
\usepackage{multicol}
\usepackage{titlesec}
\usepackage{lmodern}
\usepackage[T1]{fontenc}
\usepackage{microtype}

\newcommand{\hdr}[1]{\noindent\textbf{#1}\par\vspace{0.4em}}
\newcommand{\spacer}{\vspace{0.75em}}
\newcolumntype{Y}{>{\raggedright\arraybackslash}X}
\setlist[itemize]{leftmargin=1.1em}

\usepackage{draftwatermark}
\SetWatermarkLightness{0.95}
\SetWatermarkText{\tt DRAFT - TZstats}
\SetWatermarkScale{0.5}

\usepackage{xcolor}

\usepackage[authoryear]{natbib}

\begin{document}

\title{AI Education in Higher Education: A Taxonomy for Curriculum Reform and the Mission of Knowledge}

\author{Tian Zheng (tian.zheng@columbia.edu)\\
       \small Department of Statistics,
       Columbia University\\
       \small 1255 Amsterdam Avenue, New York, NY 10027 
       }
\date{\small  Version: \today}
       
\maketitle

\begin{abstract}
Artificial intelligence (AI) is reshaping higher education, yet current debates often feel tangled, mixing concerns about pedagogy, operations, curriculum, and the future of work without a shared framework. This paper offers a first attempt at a taxonomy to organize the diverse narratives of AI education and to inform discipline-based curricular discussions. We place these narratives within the enduring responsibility of higher education: the mission of knowledge. This mission includes not only the preservation and advancement of disciplinary expertise, but also the cultivation of skills and wisdom, i.e., forms of meta-knowledge that encompass judgment, ethics, and social responsibility. For the purpose of this paper's discussion, AI is defined as adaptive, data-driven systems that automate analysis, modeling, and decision-making, highlighting its dual role as enabler and disruptor across disciplines. We argue that the most consequential challenges lie at the level of curriculum and disciplinary purpose, where AI accelerates inquiry but also unsettles expertise and identity. We show how disciplines evolve through the interplay of research, curriculum, pedagogy, and faculty expertise, and why curricular reform is the central lever for meaningful change. Pedagogical innovation offers a strategic and accessible entry point, providing actionable steps that help faculty and students build the expertise needed to engage in deeper curricular rethinking and disciplinary renewal. Within this framing, we suggest that meaningful reform can move forward through structured faculty journeys: from AI literacy to pedagogy, curriculum design, and research integration. The key is to align these journeys with the mission of knowledge, turning the disruptive pressures of AI into opportunities for disciplines to sustain expertise, advance inquiry, and serve society.
\end{abstract}

\newpage \tableofcontents

\newpage 
  \noindent \includegraphics[width=\textwidth]{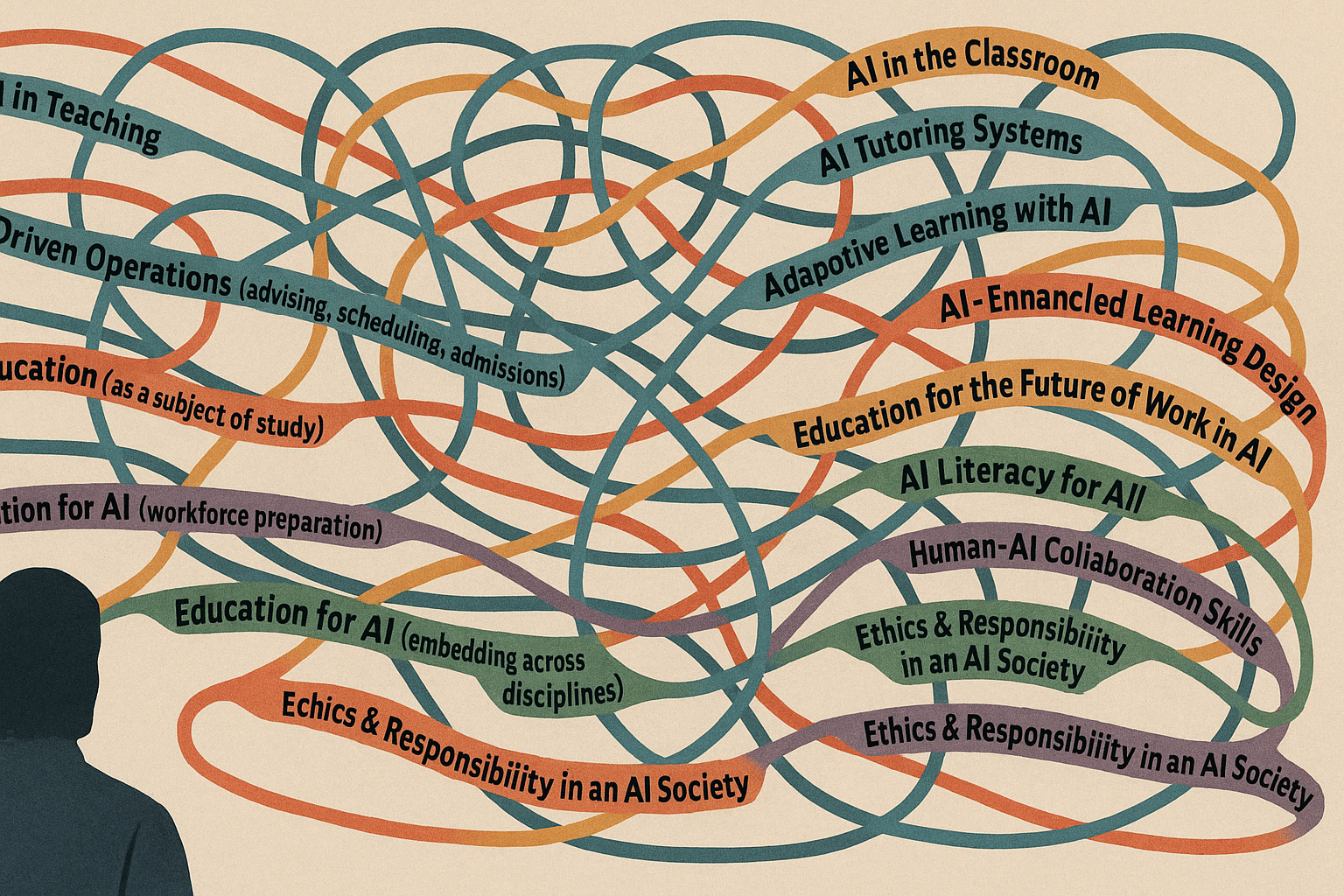}
  \textit{Image credit: ChatGPT5}

\section{Introduction}

Narratives about artificial intelligence (AI) in higher education are multiplying, with overlapping but uneven focus on pedagogy, operations, curriculum, research, and the future of work \citep{lau_ban_2023, watson_leading_2025}. The implications in each domain are profound yet distinct, and when discussed together without clarity, the result is a landscape that feels tangled and confusing \citep{ahmadOpinionWhatAI2025}. This confusion risks paralyzing institutions, faculty, and students at precisely the moment when intentional action is needed \citep{mah_artificial_2024,  mowreader_data_2025,silva_university_2025}.

This paper develops a coherent taxonomy for understanding AI in higher education by connecting these narratives to the enduring responsibility of higher education: the mission of knowledge. We begin in Section 2 by defining AI and examining its dual role as enabler and disruptor across disciplines \citep{AIFutureEducation}, showing how its hallmarks of automation, adaptivity, and integration both accelerate inquiry and unsettle expertise. We then revisit the mission of knowledge in higher education (Section 3), cultivating expertise, judgment, creativity, and responsibility, and argue that AI poses both opportunities and threats to this mission, with consequences for the future of work and society. In this framing, the mission of knowledge also encompasses the cultivation of skills and what may be referred to as “wisdom” \citep{carvalhoEducation2070Conceptual}, i.e., meta-knowledge about ethics, judgment, and social responsibility. These dimensions are as central as technical expertise, and they highlight why the discussion is not only about disciplines adopting data and machine learning, but also about preparing students and faculty to exercise good judgment in an AI-rich society.

Building on this mission-level perspective, we analyze why campus debates about AI feel so tangled in Section 4. Common phrases such as “AI in the classroom,” “AI curriculum,” and “AI for the future of work” point to different tiers of concern: practice (pedagogy), system (operations), curriculum (knowledge and disciplines), and purpose (societal futures). We show how supporting arguments often emphasize pedagogy, while opposing arguments appeal to curricular integrity, and why lasting change requires linking the “how” of teaching with the “what” of expertise. 

We then turn to disciplines as the engines of change in Section 5. Disciplines evolve when research, curriculum, and pedagogy cohere; curricular reform institutionalizes advances, pedagogy animates knowledge and drives adaptation, and faculty expertise grows through this interplay \citep{temper_higher_2025}. In the age of AI, aligning disciplinary reform with higher education’s mission of knowledge is critical for sustaining the integrity and relevance of knowledge in society.

Finally, in Section 6, we propose a way forward: curricular reform through meaningful steps and journeys. We emphasize that while pedagogy and operations (Tiers 1 and 2, as explained in Section 4) provide more immediate and resource-supported entry points, the deeper challenges of curriculum and disciplinary purpose (Tiers 3 and 4) are where lasting impact lies. Strategic pedagogical experiments are valuable not as ends in themselves, but as stepping stones that help faculty develop the expertise and confidence needed to engage with the more consequential work of curricular reform and disciplinary renewal. We need clear and supported pathways to act, moving from AI literacy to pedagogical adaptation, curriculum design, and ultimately research integration. These pathways depend on institutional structures that provide time, resources, and recognition, ensuring faculty are empowered and enabled as education leaders and disciplinary innovators. Students can be engaged as co-designers and co-developers in this process, creating feedback loops between teaching and inquiry. By focusing on the mission of knowledge and making it a collective journey, higher education can transform the overwhelming challenge of AI into a generative opportunity, ensuring that AI becomes not just a tool of efficiency or disruption, but a catalyst for the renewal of the highest purposes of education.

\section{AI: Definition and Disruption Across Disciplines}

\subsection{Defining AI}
In this discussion, we define \textbf{artificial intelligence (AI)} as: 

\begin{quote}
\emph{Computational systems that combine machine learning, statistics, optimization, and analytics to build adaptive, data-driven algorithms that automate aspects of analysis, modeling, and decision-making. Distinct from traditional data science, AI emphasizes automation, adaptivity, and the capacity to act in complex environments.}
\end{quote}

The very hallmarks that make AI powerful---automation, adaptivity, and integration into larger systems---also make it dangerous. AI’s strength lies in its ability to harness vast data and powerful models to generate insights and drive action \cite{russellArtificialIntelligenceModern1995,jordanMachineLearningTrends2015}. Its danger lies in the often poorly understood stochastic nature of data and algorithms, and the speed with which their consequences diffuse through real systems \citep{laiScienceHumanAIDecision2023}.

\subsection{AI as Enabler and Disruptor}
AI is both an \emph{enabler} of innovation and a \emph{disruptor} of established practices.

\paragraph{As Enabler.}
\begin{enumerate}
    \item \textbf{Research:} Accelerates inquiry across fields through text mining, image analysis, simulation, and predictive modeling \citep{wang_scientific_2023}.
    \item \textbf{Teaching:} Expands instructional practice via grading assistants, automated feedback, tutoring systems, and personalized learning environments \citep{saunders_pedagogy_2024,AIFutureEducation}.
    \item \textbf{Operations:} Enhances efficiency in advising, scheduling, enrollment management, and institutional analytics \citep{team_ai-ready_2025}.
\end{enumerate}

\paragraph{As Disruptor.}
AI challenges disciplines in distinct ways:
\begin{enumerate}
    \item \textbf{Tension between automation and expertise.} AI makes core tasks easier, but deep expertise remains essential to interpret and govern results. Examples: computer science, law, medicine, engineering \citep[e.g.,][]{surdenMachineLearningLaw2014,topolHighperformanceMedicineConvergence2019}.
    \item \textbf{Students adopt faster than faculty.} Learners embrace generative and analytic tools ahead of formal teaching practices, unsettling classroom authority. Examples: business, communication and media, languages, education \citep[e.g.,][]{horowitzAdoptingAIHow2024,walton_family_foundation_gen_2025}.
    \item \textbf{Subject matter transformation.} The very objects of study are reshaped by AI (e.g., political discourse, economic models, social behavior). Examples: political science, economics, sociology, history, philosophy \citep[e.g.,][]{laiScienceHumanAIDecision2023}.
    \item \textbf{Continuous methodological renewal.} Fields where methods themselves are redefined by AI, creating constant need for adaptation. Examples: AI/CS research, statistics, law and policy, ethics, environmental and health sciences \citep[e.g.,][]{grossmann_ai_2023}.

\end{enumerate}




Therefore, AI disruption does not affect all fields uniformly. While AI universally enables new forms of research, teaching, and operations, its disruptive impact depends on the relationship between automation, expertise, subject matter, and methodology within each discipline.

\subsection{Example: Statistics: the AI-Triggered ``Soul-Searching''}

Statistics provides a clear example of how AI simultaneously enables new opportunities, introduces disruptive challenges, and prompts deeper reflection on disciplinary identity \citep{lin2025statistics}.

\paragraph{Opportunities (AI as Enabler).}
AI builds directly on statistical methods, extending them into adaptive and automated systems. For statisticians, this creates new opportunities to:

\begin{itemize}
    \item Accelerate data analysis and discovery across scientific fields through automated pipelines and scalable computation.
    \item Expand teaching practice by linking foundational concepts (e.g., inference, uncertainty quantification) with applied AI systems that students encounter in real life.
    \item Reassert the discipline’s role as a source of rigor in areas such as evaluation, validation, uncertainty, and responsible use of data.
\end{itemize}

\paragraph{Challenges (AI as Disruptor).}
At the same time, AI disrupts statistics in several ways:
\begin{itemize}
    \item Core practices once central to training (manual computation, routine model fitting) can now be automated, raising questions about what should remain essential in the curriculum.
    \item Students often adopt AI tools faster than faculty can adapt instruction, leading to a mismatch between classroom exercises and real-world practice.
    \item The boundary between “statistics” and “machine learning” blurs, creating confusion in both industry and academia about where expertise lies.
\end{itemize}

\paragraph{Soul-Searching (Disciplinary Identity).}
These opportunities and challenges provoke reflection about what statistics should stand for in an AI-rich world:
\begin{itemize}
    \item Is the core of statistics its techniques, or its epistemic stance on uncertainty, variation, and inference?
    \item Should curricula emphasize computation and algorithms, or focus more deeply on interpretation, communication, and responsible use?
    \item How should statistics articulate its identity alongside computer science, data science, and AI, while maintaining its unique contributions to knowledge?
\end{itemize}

In this way, statistics illustrates the broader dilemma facing many disciplines. AI expands the reach of the field, disrupts traditional training and practice, and forces a reconsideration of what constitutes its enduring knowledge and values.

\section{Higher Education's \textit{Mission of Knowledge} in the Age of AI}

Education has never had a single, universal mission. Across time and place, it has been expected to cultivate individuals, sustain societies, fuel economies, transmit culture, and shape ethical values \citep{kerrUsesUniversityFifth2001}. These purposes coexist in tension, and that tension is not a weakness but the very condition that sustains the richness of human knowledge and the multidimensionality of persons and communities across generations \citep{Delors1996Learning}.

From this perspective, higher education occupies a distinctive role. It is not merely another stage in the pipeline of credentials, nor simply a service to the labor market. Universities can be understood as sites where knowledge lives, interacts, and evolves \citep{Newman1852TheIdea}. They preserve, interrogate, and transmit inherited knowledge; generate new forms of knowledge through research and critique; and translate knowledge into technologies, policies, and practices that circulate back into society. In this sense, higher education serves as a crucial node in the ecosystem of knowledge, binding together individuals, institutions, and the larger social fabric.

Students are perhaps the most powerful way universities return knowledge to society. They carry with them not only disciplinary expertise but also habits of inquiry, ethical orientations, and capacities for innovation. This includes critical skills and wisdom, as forms of meta-knowledge, enabling individuals to apply their learning responsibly and adaptively \citep{carvalhoEducation2070Conceptual}. Higher education must foster students' abilities to navigate ethical complexities, sustain social trust, and exercise judgment. These capacities have been made more essential by AI. It is not simply about securing employment, though economic opportunity matters. It is also about what kinds of jobs, what forms of expertise and judgment they embody, and how those roles contribute to both individual fulfillment and collective well-being. Higher education’s responsibility is to help students see their professional pathways as inseparable from the broader ecosystem of knowledge and its social purposes \citep{AIFutureEducation}.

\subsection{AI as a Challenge to the Mission}
Artificial intelligence introduces both opportunities and threats to this mission. On the one hand, AI promises to expand access, accelerate discovery, and personalize learning \citep{carvalhoEducation2070Conceptual}. On the other hand, it risks undermining the very practices by which knowledge is cultivated: the slow work of reasoning, critique, and the apprenticeship into disciplinary traditions \citep[e.g.,][]{kosmyna_your_2025}. If foundational tasks of expertise are automated without careful judgment, students may lose not only the practice but also the understanding of how knowledge is made. The threat is not simply to teaching routines or institutional operations, but to higher education’s larger role in sustaining the integrity, plurality, and social purpose of knowledge. Because knowledge is the foundation of work and society, any erosion of this mission carries consequences far beyond the university.

\subsection{Response to AI Needs Grounding in the Mission}
This framing clarifies why grounding present efforts in higher education’s mission is essential. Institutions, faculty, and students alike feel the urgency to act in response to AI, but without a shared vision of what education is for, the impulse to innovate easily fragments into disconnected experiments or defensive prohibitions \citep{mah_artificial_2024,watson_leading_2025,ahmadOpinionWhatAI2025}. The most effective path forward requires situating decisions about AI—whether in pedagogy, curriculum, or institutional operations—within a clear understanding of education’s purpose: cultivating expertise, judgment, creativity, and responsibility for the future of knowledge. 

\subsection{Faculty, Students, and Institutions as Agents}
Responding to AI is not only a matter of adopting tools; it is about strengthening the agency of the core participants in the knowledge ecosystem:
\begin{itemize}
    \item \textbf{Faculty} are not only instructors but also knowledge makers whose pedagogy and research evolve together. Their growth as educators and scholars is central to aligning AI with educational values \citep{mollick_instructors_2024,mah_artificial_2024}.
    \item \textbf{Students} are not passive recipients but emerging participants in the life of knowledge. Engaging with AI should help them practice inquiry, judgment, and application, not bypass these capacities \citep{ahmadi_student_2023,zeivots_co-design_2025}.
    \item \textbf{Institutions} provide the structures—curricula, policies, and communities—that enable faculty and students to enact the mission \citep{polly_examining_2021,temper_higher_2025,watson_leading_2025}. Their challenge is to coordinate innovation so that individual efforts scale toward shared goals rather than scatter into isolated responses.
\end{itemize}

Taken together, this ``Mission of Knowledge'' perspective positions higher education not as a narrow economic instrument but as a vital steward of knowledge in society. In the age of AI, this enduring mission must be the anchor from which we decide how technology should reshape learning and teaching: as a tool that supports, but never replaces, the fundamental work of education. Grounding our responses in this mission ensures that curricular rethinking, pedagogical adaptation, and faculty and student agency remain aligned in guiding change at the scale this moment demands.

Also in this framing, higher education needs to respond proactively to AI and clarify its \textit{mission of knowledge} in the age of AI. Yet acting is difficult. Limited resources are certainly a root constraint, but institutions often encounter confusion and fragmentation well before they hit the wall of resources. As the next section shows, debates about AI in education are tangled across operations, pedagogy, curriculum, and disciplinary purpose. Without a clear mission-level vision to orient these dimensions, institutions, faculty, and students struggle to translate urgency into coherent action.

\section{Why ``AI Education'' Narratives
Feel Tangled and Confusing}

The discourse on artificial intelligence (AI) in higher education is marked by overlapping narratives and conflicting sentiments. On many campuses, conversations about AI are animated by a mixture of sentiments \citep{mah_artificial_2024,watson_leading_2025}.

\begin{itemize}
    \item \textbf{Excitement and eagerness}: a sense of opportunity to modernize teaching, research, and institutional practices.
    \item \textbf{Wariness or resistance}: skepticism about unproven claims, cultural inertia, or attachment to existing practices.
    \item \textbf{Uncertainty and confusion}: difficulty in determining what exactly to do, how to implement change, and what should be prioritized.
    \item \textbf{Anxiety and threat}: fears of being displaced or rendered obsolete, without clear guidance on how to prepare.
    \item \textbf{Contradictory practices}: the tension between institutions adopting AI to optimize operations while simultaneously banning it in classrooms under academic integrity policies.
\end{itemize}

These sentiments coexist because ``AI in education'' is not a single narrative but a bundle of overlapping conversations. Some focus on teaching practices, some on learning outcomes, others on disciplinary identity, or on the mission of higher education itself. When discussed together without clarification, these distinct narratives create the impression of confusion.

\subsection{Common AI Education Phrases}

\subsection*{Common AI-in-Education Phrases by Tier}

\begin{itemize}
  \item \textbf{Tier 1: Point-of-Use (Practice-Level)} \\
  Direct applications in classroom teaching and learning.
  \begin{itemize}
    \item AI in Teaching
    \item AI in the Classroom
    \item AI Pedagogy
    \item AI Tutoring Systems
    \item Adaptive Learning with AI
  \end{itemize}

  \item \textbf{Tier 2: Design \& Systems (Institution-Level)} \\
  How institutions structure curriculum, pedagogy, and operations.
  \begin{itemize}
    \item AI-Supported Instruction
    \item AI-Enhanced Learning Design
    \item AI in Higher Education
    \item AI-Driven Operations (advising, scheduling, admissions)
    \item AI for Student Support
    \item AI \& Learning Environments
  \end{itemize}

  \item \textbf{Tier 3: Curriculum \& Knowledge (Disciplinary-Level)} \\
  How knowledge itself shifts, and what is taught.
  \begin{itemize}
    \item AI Education (as a subject of study)
    \item Education for AI (workforce preparation)
    \item Education with AI (embedding across disciplines)
    \item Curriculum Reform for the Age of AI
  \end{itemize}

  \item \textbf{Tier 4: Societal Futures (Purpose-Level)} \\
  How education prepares learners for broader economic and societal transformations.
  \begin{itemize}
    \item Education for the Future of Work in AI
    \item AI Literacy for All
    \item Human--AI Collaboration Skills
    \item Ethics \& Responsibility in an AI Society
  \end{itemize}
\end{itemize}

\subsection{Dimensions of AI Education Narratives}

This proliferation of terms contributes to the sense of confusion on campus. Many of them overlap or blur across tiers, yet each points to a different dimension of AI’s impact on higher education. For clarity, a streamlined lexicon can help: 

\begin{itemize}
  \item \textbf{AI Pedagogy} (Tier 1): methods of teaching with or around AI. This is the most immediate level: AI as tutoring systems, writing assistants, grading tools, or sources of student anxiety. The gains are real — more timely feedback, new modes of practice, potential for personalized learning — but so are the risks of over-reliance and integrity erosion. Faculty need literacy and support to make informed pedagogical choices.

  \item \textbf{AI in Higher Education Operation} (Tier 2): systemic adoption, operations, and institutional design. 

  \item \textbf{AI Curriculum} (Tier 3): both \emph{education for AI} (training specialists) and \emph{education with AI} (embedding across disciplines). Here, AI is both subject and tool. Some fields face tensions as AI makes certain tasks easier (coding, diagnostics) but still demand deep expertise. Others find students adopting AI faster than faculty can adapt. Still others see their subject matter transformed entirely (political science, law, history, philosophy). This tier is where reform sparks new research — curriculum becomes the medium that links education with inquiry.

  \item \textbf{AI \& the Future of Work and Society} (Tier 4): preparing graduates for AI-shaped labor markets and societal roles. Finally, AI raises big questions about how universities prepare students for work and citizenship. It is not only about jobs but about what kinds of jobs, and how those roles contribute both to personal careers and to society. This tier is where ethics, justice, and responsibility must be embedded, so that AI serves humanity rather than undermining it.
  
\end{itemize}

\noindent This lexicon provides a framework to treat these categories as \emph{dimensions of AI education narratives} that must be disentangled for more productive institutional dialogue. For example, in Table~\ref{tab:ai-narratives}, we can see clearly where excitement, anxiety, and tension concerning AI education fall. 
\begin{table}[htbp]
\centering \small
\caption{Four dimensions of AI education narratives and the sentiments they evoke.} \vspace{1em}
\renewcommand{\arraystretch}{1.25} 
\label{tab:ai-narratives}
\small
\begin{tabular}{p{5.3cm}p{10.5cm}}
\textbf{Dimension} & \textbf{Common Sentiments} \\
\hline

\textbf{Pedagogy (Tier 1)} & Confusion about appropriate classroom rules; tension between enhancing learning with AI vs.~academic integrity concerns; wariness about erosion of traditional teaching roles. \\
\hline
\textbf{Operations (Tier 2)} & Excitement about efficiency gains (admissions, advising, grading, scheduling); anxiety about surveillance, loss of human connection, or dehumanization. \\
\hline
\textbf{Curriculum (Tier 3)} & Broad consensus that AI literacy is essential for future work; disagreement on what counts as core vs.~peripheral training; anxiety about automation of once-essential skills. \\
\hline
\textbf{Societal Relevance (Tier 4)} & Anxiety about erosion of disciplinary expertise; eagerness to claim relevance; uncertainty about long-term roles in an AI-shaped society. \\
\hline
\end{tabular}
\end{table}
The coexistence of these dimensions explains the ``tangled'' feel of AI discussions. Nearly everyone agrees that students must be prepared for a future where AI is central, but permitting AI use in the classroom does not by itself achieve that goal. For meaningful and scalable impact, discussions at the level of teaching practice and classroom integration (Tiers~1 and~2) need to be anchored in a clearer vision at the level of curriculum and societal purpose (Tiers~3 and~4). These higher-level tiers define the educational values and goals that classroom and operational practices should serve. At present, such re-visioning is emerging unevenly across disciplines, which further complicates the landscape. This unevenness also helps explain why faculty, students, and institutions alike often feel overwhelmed: there is a strong and urgent impulse to act, yet the absence of a shared curricular and disciplinary vision leaves stakeholders unsure how to reconcile immediate pedagogical choices with long-term educational purposes. 

\subsection{Curricular Questions in the AI Era}
A productive way to think about curricular impact is through an exercise: within every discipline’s curriculum, some practices are deliberately tedious or repetitive. They serve both as training exercises and as thresholds of expertise. AI raises critical questions:

\begin{itemize}
    \item Are these \emph{tedious tasks} essential for cultivating core expertise, or are they bottlenecks that can be responsibly automated?
    \item If AI automates them, does the discipline lose something central to its subject identity, or does it merely accelerate students toward higher-order work?
    \item If AI frees up bandwidth, what new problems become possible or even urgent to pursue? What learning outcomes should be elevated in their place?
\end{itemize}

All of these are curricular questions, because they concern the \emph{design of the training sequence that defines expertise}. Yet current debates on AI in education often conflate pedagogy (``how to teach with AI'') with content (``how to teach AI as expertise''). For example, supporting arguments for \textit{AI in education} typically emphasize pedagogy and classroom practice (how to teach \textit{better} with AI), while opposing arguments frequently invoke curricular integrity and disciplinary value (what knowledge and expertise should define the field going forward). When treated out of context, these perspectives talk past one another in many disciplines.  In reality, questions of how to teach with AI cannot be separated from what students need to learn for an AI-rich future. Together, they point to the more fundamental challenge: \emph{how to reimagine the curriculum itself to be AI-native}.

\subsection{What Needs to Happen}
Moving forward, the key challenge is not simply distinguishing between pedagogy, curriculum, and disciplinary identity, but understanding how they work together in shaping how higher education responds to AI. Curriculum rethinking provides the vision for what expertise and values a discipline should advance; curricular reform translates that vision into programs of study; pedagogy animates the reform through new practices of teaching and learning; and faculty expertise grows in the process of adapting to and guiding these changes. Students, in turn, encounter the evolving discipline through this interplay. At present, the easier work is happening in Tiers 1 and 2, where resources and cross-disciplinary examples are more readily available. Yet if these efforts are not guided by, and ultimately embedded in, a vision for curriculum and disciplinary purpose, they risk remaining fragmented. Starting with pedagogy is therefore a strategic move, but only if it is explicitly connected to deeper Tier 3 and 4 discussions. The next section unpacks these relationships in greater detail, showing how they form the engine of disciplinary change in an AI-rich era.

\section{Discipline Advances Through
Curricular Reform, Driven by Pedagogical Practices and Development of Faculty Expertise}

To offer a way to address the tension between the pressing need to respond and the lack of shared clarity, this section turns to the interlocking roles of curriculum, pedagogy, and faculty expertise as engines of meaningful change.

Here we offer a conceptual analysis of what a 
\emph{discipline} is and how it evolves. It argues that curricular change is the principal mechanism through which a discipline institutionalizes intellectual and societal shifts, while pedagogy---distinct from curriculum---both animates disciplinary knowledge and can actively drive curricular reform. We further highlight how research- and practice-integrated pedagogy functions as a form of faculty development that feeds back into disciplinary transformation.

\subsection{What Is a Discipline?}
A discipline is more than a topic area; it is a socially organized field of knowledge and practice. It is recognizable through five interlocking features:
\begin{enumerate}
  \item \textbf{Domain of inquiry}: a shared set of phenomena or problems deemed legitimate for study (e.g., physical laws in physics, social interaction in sociology).
  \item \textbf{Conceptual and methodological framework}: the paradigms, theories, methods, exemplars, and standards of evidence that orient inquiry.
  \item \textbf{Community of practitioners}: scholars and professionals who uphold and reproduce norms through peer review, advising, conferences, and journals.
  \item \textbf{Institutional structures}: departments, curricula, professional societies, accreditation regimes, and funding streams.
  \item \textbf{Boundary-setting}: distinctions between what counts as central, peripheral, or outside the field; judgments about legitimate questions and techniques.
\end{enumerate}
Crucially, a discipline persists by \emph{codifying, teaching, and reproducing itself} across generations. This is where curriculum and pedagogy become constitutive, rather than incidental.

\subsection{How Disciplines Evolve}
Disciplines change under pressure from three broad sources, two are drivers, one is a lever:
\begin{itemize}
  \item \textbf{Driver I: Intellectual advances}: new theories, methods, instruments, and data reshape what is thinkable and doable.
  \item \textbf{Driver II: Societal demands}: public needs and ethical concerns (e.g., climate risk, data governance, AI safety) reorder priorities.
  \item \textbf{Lever: Institutional Incentives}: accreditation, funding, and policy shifts alter what is rewarded or required.
\end{itemize}
Yet frontier research alone does not secure durable change. Disciplinary evolution is not complete until new ideas are \emph{absorbed into curriculum}---translated into teachable sequences, competencies, and assessments that train future practitioners. Without curricular uptake, innovations remain isolated at the frontier and fail to reshape the field’s mainstream identity.

\subsection{What Curriculum Means to a Discipline?}
\textit{Curriculum} is the translation of disciplinary knowledge into learnable form. It specifies the content to be taught, its scope and sequence, and the standards by which mastery is judged. Through curriculum, disciplines:

\begin{enumerate}
  \item \textbf{Codify identity}: articulating what is foundational versus advanced, and what falls outside the canon.
  \item \textbf{Reproduce membership}: onboarding novices into competent participation through prerequisite chains, labs, studios, and capstones.
  \item \textbf{Adapt to change}: incorporating new subfields, methods, and values (e.g., data ethics, reproducibility practices).
  \item \textbf{Legitimize knowledge}: inclusion signals centrality; exclusion risks marginalization.
\end{enumerate}
Because curriculum determines what every graduate is expected to know and be able to do, it is the discipline’s most powerful lever for self-definition and renewal.

\subsection{Curricular Change as the Engine of Disciplinary Shift.} When a discipline updates its curriculum, it effectively redefines itself for the next generation. Canonical examples include the addition of statistics to psychology majors, design and teamwork to engineering, or ethics and data governance to computer science. Such reforms:
\begin{itemize}
  \item Reframe the \textbf{problem space} students are trained to notice and prioritize.
  \item Normalize \textbf{methods and tools} (e.g., simulation, fieldwork, reproducible workflows).
  \item Align \textbf{assessment} with new competencies (e.g., portfolios, open-ended projects, replication assignments).
\end{itemize}
Because graduates carry these repertoires into research and practice, curricular change compounds into disciplinary change: over time it alters the questions asked, the evidence considered persuasive, and the communities with which the discipline interacts.

\subsection{How Pedagogy Differs from Curriculum?}
\textit{Pedagogy} concerns how learning happens: the philosophies, strategies, and practices of teaching. Lectures, seminars, inquiry- and problem-based learning, design studios, role-play, and case-based teaching are pedagogical choices. The distinction is succinct:
\begin{quote}
\textbf{Curriculum = what is taught. \quad\quad\quad Pedagogy = how it is taught.}
\end{quote}
They are separable: a department may add machine learning to the curriculum yet teach it via traditional lectures; conversely, a long-standing algorithms course can be taught through active learning without changing the topic list. The two dimensions are, however, interdependent in their effects on learning and identity.

\subsection{How Pedagogy Drives Curricular Reform}
Although curricular reform often leads, pedagogy can be a \emph{driver} of change:
\begin{enumerate}
  \item \textbf{Surfacing teachability}: Visual or block-based environments revealed that core computing concepts are accessible to novices, prompting a shift from syntax-first to concepts-first introductory sequences.
  \item \textbf{Redefining relevance}: Project-based and community-engaged learning require authentic contexts, forcing curricula to include teamwork, stakeholder analysis, and ethical reflection.
  \item \textbf{Exposing blind spots}: Case-based pedagogy in AI and data courses surfaces fairness, accountability, and social impact, catalyzing formal ethics modules and cross-disciplinary collaborations.
  \item \textbf{Promoting inclusion}: Culturally responsive and inquiry-driven teaching reveals structural barriers (assumed prior knowledge, narrow exemplars), pushing curricula to broaden entry points and diversify cases.
\end{enumerate}
Here pedagogy functions as a \emph{probe}: by altering the mode of engagement, it reveals gaps in content, sequencing, or prerequisite structures, compelling curriculum to adapt. In this sense, early pedagogical shifts (Tier 1) are not endpoints but entry points. They provide faculty with the chance to test practices, surface new questions, and build the shared acumen needed to inform deeper Tier 3 and Tier 4 reforms in curriculum and disciplinary purpose. Starting with pedagogy is therefore a strategic move: it is accessible and supported by available tools, but it must be consciously linked to longer-term curricular rethinking.

\subsection{Pedagogy as Faculty Development}
Pedagogy shapes not only student learning but also faculty expertise. Research- and practice-integrated pedagogies (e.g., studio, capstone, inquiry) require instructors to:
\begin{itemize}
  \item Update knowledge of current tools, datasets, and professional workflows to design authentic tasks.
  \item Translate cutting-edge research into teachable modules, clarifying conceptual cores and methodological trade-offs.
  \item Work as co-inquirers with students, yielding fresh perspectives and sometimes seeding new research.
  \item Expand professional identity from \enquote{content gatekeeper} to \enquote{mentor and disciplinary steward.}
\end{itemize}
In this sense, pedagogy operates as a continuous professional learning mechanism. As faculty grow through teaching, they become credible advocates for curricular reforms that better align with contemporary practice and values. The feedback loop is clear: \emph{innovative pedagogy $\rightarrow$ faculty learning $\rightarrow$ curricular change $\rightarrow$ disciplinary evolution}.

This dynamic illustrates why Tier 1 and Tier 2 efforts are valuable even if they seem modest compared to the \textit{larger questions of curriculum and disciplinary identity}. By experimenting with pedagogy and operations, faculty are not only supporting students in the present but also training themselves to engage with the more complex work of defining what expertise, knowledge, and purpose their discipline should uphold in the age of AI. In other words, \textit{what faculty need to teach is also what they need to learn}: pedagogical adaptation becomes the pathway into curricular reform and disciplinary renewal.

\subsection{Example: Curricular Reform in Computer Science}

Computer science demonstrates how curricular content, pedagogy, and faculty expertise co-evolve. In earlier decades, core PhD training emphasized foundational systems areas such as operating systems and compilers. Today, many programs treat these as electives, while expanding core requirements to include machine learning, data science, and ethics. These curricular shifts have been accompanied by pedagogical innovation—project-based teamwork for software engineering, block-based environments for novices, and case-based methods for ethics. Faculty development has been central throughout, as instructors retool to teach new areas while sustaining disciplinary identity. Computer science thus illustrates how disciplines adapt by renegotiating both what is considered “core” and how that knowledge is taught.

\subsection{How To Move Forward for AI Education}
Disciplines are living systems that evolve through the mutual shaping of research, curriculum, and pedagogy. Research expands the frontier; curriculum institutionalizes advances; pedagogy animates knowledge and often reveals what the curriculum must become. Effective governance of a discipline therefore requires attention to all three: incentives for intellectual innovation, structures for curricular renewal, and support for pedagogical experimentation that doubles as faculty development. When these elements cohere, the discipline remains vibrant, accessible, and socially responsive.

In the age of AI, this co-evolutionary process must be understood in light of higher education’s broader mission of knowledge. If disciplines govern themselves well, by aligning research, curriculum, and pedagogy, they help sustain this mission by preparing students to carry forward expertise, judgment, and creativity into work and civic life. If they fail, AI risks being absorbed only as a tool of efficiency or disruption, rather than as a catalyst to renew the role of education in cultivating human knowledge. Linking disciplinary reform back to the mission of knowledge ensures that institutional responses to AI remain anchored in education’s highest purposes, while still equipping students and faculty to navigate a rapidly changing world.

In this framing, the critical questions of curricular reform and disciplinary identity are paramount, but faculty cannot reach them without first experimenting in pedagogy and operations. What they need to teach is also what they need to learn. Faculty build acumen for deeper curricular debates by starting with Tier 1 and Tier 2 practices, provided these are framed as steps toward Tier 3 and Tier 4 renewal.

\section{AI Curricular Reform Through Meaningful Steps and Journeys}

Higher education’s response to AI cannot rest on ad hoc experiments or isolated classroom practices. As discussed in the previous section, to meet the scale of the moment, reform must be grounded in the curriculum of a discipline as the central lever, where disciplines define their expertise, prepare students for future work and citizenship, and link teaching with research. Yet the mission of AI education today often feels so expansive and consequential that faculty and institutions struggle to see where to begin. Faced with this grand vision, many experience paralysis: a strong impulse to act collides with uncertainty about where to start. What is needed are meaningful journeys—structured, supported, and faculty-led—that translate the mission of knowledge in the age of AI into concrete and feasible steps, given existing resources and stakeholder interests. These journeys begin most feasibly in pedagogy and operations (Tiers 1 and 2), where examples and resources are easier to access and cross-disciplinary learning is possible. But they must always be guided by a vision of curriculum and disciplinary purpose (Tiers 3 and 4), which take longer to develop yet ultimately matter most for sustaining expertise and identity.

\subsection{Faculty Development and Institutional Support}
Faculty are at the heart of curriculum design, but they cannot and should not carry this work alone. Institutions bear responsibility for creating the structures—resources, training, and recognition—that make faculty-led curricular reform possible. 
Part of today’s challenge is that the scope and weight of AI education can feel urgent, overwhelming, and uncertain. Many are not sure how to move from broad aspirations to concrete steps. Institutions can play a catalytic role by:
\begin{itemize}
    \item Establishing \textbf{faculty development programs} that provide time, resources and training to explore the implications of AI for their disciplines.
    \item Creating \textbf{centralized hubs} that curate tools, frameworks, and examples while coordinating interdisciplinary exchange.
    \item Incentivizing \textbf{curriculum reform projects} that allow faculty to experiment and iterate with institutional backing.
\end{itemize}

\subsection{Research–Education Integration}
A distinctive opportunity lies in integrating research and education so that curricular reform is not only a teaching initiative but also a mode of scholarly inquiry. Faculty can invite energetic students as co-designers and co-developers of new courses, modules, and learning resources. In doing so, students are positioned not only as learners but also as partners in the knowledge ecosystem, accelerating adaptation and linking classroom reform to the broader mission of returning knowledge to society. This model:
\begin{itemize}
    \item Engages students in shaping the future of their disciplines.
    \item Bridges research and teaching by embedding inquiry into curricular design.
    \item Multiplies capacity for innovation while grounding reform in lived classroom experience.
\end{itemize}
By linking education reform with research, institutions create feedback loops that both sustain disciplinary renewal and generate new scholarly directions.

\subsection{Staged Faculty Journeys in Curricular Reform}
For reform to be scalable, faculty need clear and practical pathways for adoption. We propose a staged progression grounded in the vision of curricular reform and disciplinary renewal as the ultimate goals. These stages are not meant to be followed in a rigid sequence, nor are they a checklist of obligations. Instead, they provide a flexible map that allows faculty, supported by their institutions and energized by their students, to take timely steps with available resources while still aiming toward long-term renewal.

\begin{enumerate}
    \item \textbf{AI Literacy:} Developing a baseline understanding of AI and its implications, including critical perspectives on ethics, bias, and limitations.
    \item \textbf{AI Pedagogy:} Exploring ways to integrate AI into classroom practices, from tutoring systems to assignment design and assessment.
    \item \textbf{AI Curriculum:} Designing courses and programs that incorporate AI as content or reshape existing curricula to be AI-informed.
    \item \textbf{Research Integration:} Linking curricular innovation to emerging research agendas, enabling faculty to feed teaching insights back into disciplinary inquiry.
\end{enumerate}

This staged journey reframes faculty development as \emph{learning by teaching}, positioning curriculum reform as the bridge into deeper research engagement. In this view, curricular reform is not just a strategy for adapting to AI but a generative pathway for re-envisioning disciplinary knowledge and practice. It brings together institutional support, faculty growth, and student participation in a shared effort, ensuring that responses to AI move beyond scattered experiments toward coordinated renewal anchored in the mission of knowledge. 

Practical worksheets (Figure~\ref{fig:journeys}) are provided in the Appendix to help departments and disciplines apply this framework in their own contexts.

\begin{figure}[ph]
  \centering
  \includegraphics[width=\linewidth]{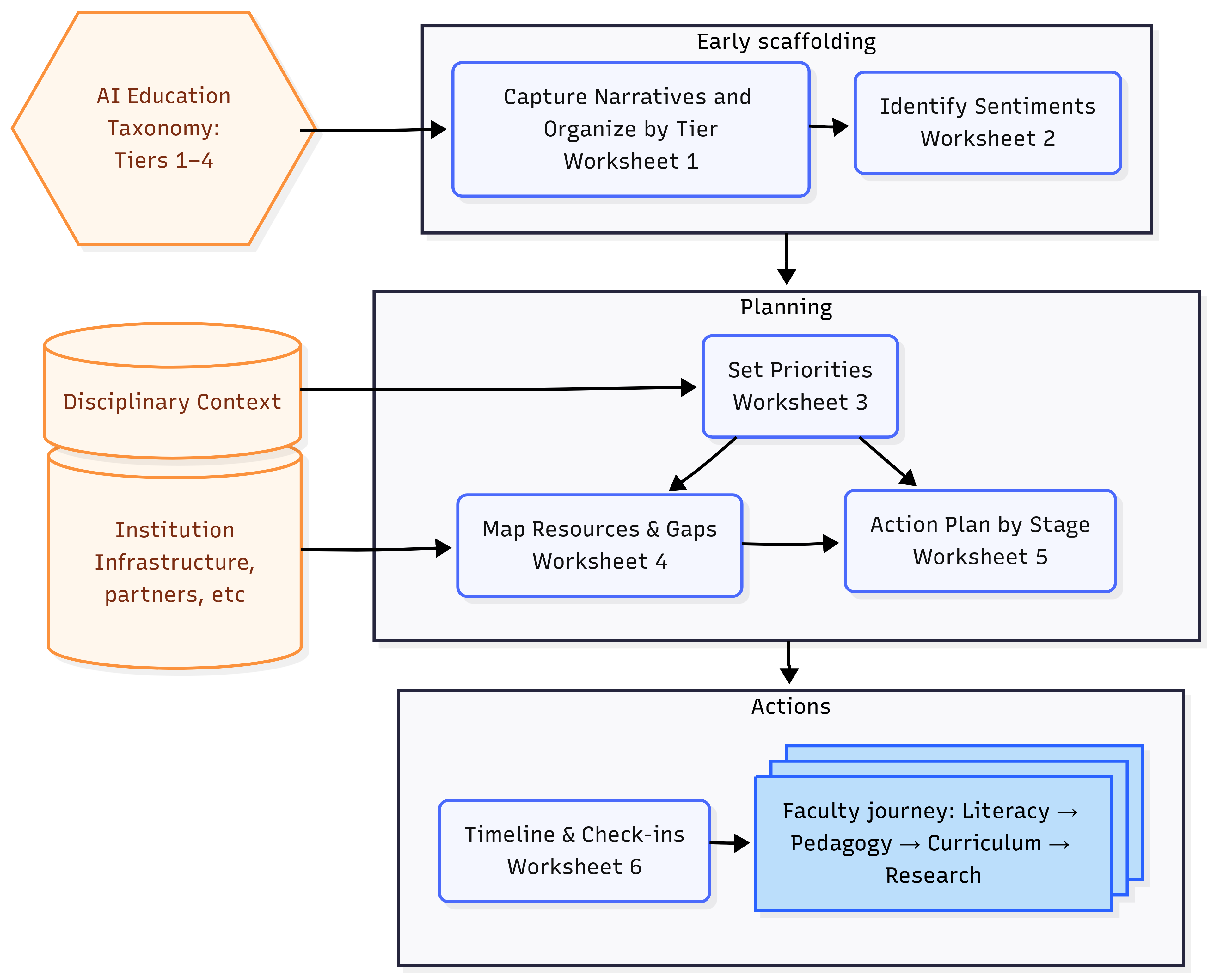}
  \caption{An example workflow for a discipline/department's AI curriculum reform. Start by capturing narratives and situating them in tiers, through identifying sentiments, priorities, and resources, to build staged action plans with timelines and regular check-ins. Early steps scaffold shared understanding and confidence (Tiers 1–2), while later steps guide deeper curricular and disciplinary renewal (Tiers 3–4). Example worksheets can be found in the Appendix.}
  \label{fig:journeys}
\end{figure}

\vskip 0.2in
\bibliographystyle{apalike}
\bibliography{reference}






\newpage
\addcontentsline{toc}{section}{Appendix: Worksheets for Mapping AI Narratives, Sentiments, Priorities, and Actions (with Examples)}

\section*{Appendix: Discipline/Department Worksheets: Mapping AI Narratives, Sentiments, Priorities, and Actions (with Examples)}

\noindent The following worksheets are intended as practical tools for departments or disciplinary groups that are seeking ways to explore thoughtfully their paths forward in response to the ``AI in higher education'' narratives.  They are aligned with the taxonomy and staged faculty journeys introduced in the main text. Their purpose is not to prescribe uniform immediate actions, but to help structure dialogue,
self-assessment, and planning. Each worksheet focuses on a different dimension: mapping
narratives, surfacing sentiments, prioritizing areas for attention, identifying resources and
gaps, and outlining staged actions. Together, they are meant to guide collective reflection and
help communities translate their own mission of knowledge, in the age of AI, into concrete, locally grounded steps. \\

\noindent \textit{Note: The entries in \textcolor{SkyBlue}{blue} text below are illustrative examples. Replace them with a discipline or department’s details.}

\subsection*{Worksheet 1. Map the Narratives: What are people actually talking about in our discipline or department?}

\hdr{Instructions:} 
\noindent List dominant AI-related narratives in your department. Tag each with a tier: Tier 1 = Practice/Pedagogy; Tier 2 = Operations/System; Tier 3 = Curriculum/Knowledge; Tier 4 = Purpose/Future of Work \& Society.\\

\renewcommand{\arraystretch}{1.3}
\noindent \begin{tabularx}{\textwidth}{@{}p{4.2cm} p{1cm} Y Y@{}}
\toprule
\textbf{Narrative (phrase)} & \textbf{Tier} & \textbf{Who voices it? (courses, roles, groups)} & \textbf{Why it matters here (implications/needs)} \\
\midrule
\textcolor{SkyBlue}{AI in the Classroom} & \textcolor{SkyBlue}{1} & \textcolor{SkyBlue}{Intro Course instructors; TAs} & \textcolor{SkyBlue}{Need assignment redesign and integrity policy alignment} \\
\textcolor{SkyBlue}{AI for Student Support (advising/chatbots)} & \textcolor{SkyBlue}{2} & \textcolor{SkyBlue}{Student success office; program directors} & \textcolor{SkyBlue}{Could scale advising; must address data/privacy and accuracy} \\
\textcolor{SkyBlue}{AI Curriculum (what to teach, where)} & \textcolor{SkyBlue}{3} & \textcolor{SkyBlue}{Curriculum committee} & \textcolor{SkyBlue}{Identify insertion points; release time for course redesign} \\
\textcolor{SkyBlue}{AI \& Future of Work} & \textcolor{SkyBlue}{4} & \textcolor{SkyBlue}{$\cdots$} & \textcolor{SkyBlue}{Align skills language;}  \\
\bottomrule
\end{tabularx}

\spacer

\newpage \subsection*{Worksheet 2. Surface the Sentiments: How do people feel about AI in Education for our discipline/department?}

\hdr{Instructions} 
\noindent Capture the range of sentiments you’re hearing. Use brief quotes or examples. Note implications for action.\spacer

\noindent
\begin{tabularx}{\textwidth}{@{}p{2.3cm} Y Y Y@{}}
\toprule
\textbf{Sentiment} & \textbf{Who expresses it} & \textbf{Example quote/behavior} & \textbf{Implications (risk/opportunity)} \\
\midrule
\textcolor{SkyBlue}{Excitement} & \textcolor{SkyBlue}{Early adopters; students} & \textcolor{SkyBlue}{“Finally, feedback at scale.”} & \textcolor{SkyBlue}{Channel into pilots; create exemplar repository} \\
\textcolor{SkyBlue}{Wariness, Resistance} & \textcolor{SkyBlue}{Senior faculty} & \textcolor{SkyBlue}{“Hype cycle—we’ve seen this before.”} & \textcolor{SkyBlue}{Analyze evidence; offer opt-in/opt-out paths}\\
\textcolor{SkyBlue}{Uncertainty, Confusion} & \textcolor{SkyBlue}{Department leadership} & \textcolor{SkyBlue}{“Where/how do we start?”} & \textcolor{SkyBlue}{Need to design staged pathways; clarify roles/resources} \\
\textcolor{SkyBlue}{Anxiety, Threat} & \textcolor{SkyBlue}{Lab-heavy courses} & \textcolor{SkyBlue}{“Will this deskill our field?”} & \textcolor{SkyBlue}{Clarify core expertise; define, justify and protect signature competencies.} \\
\textcolor{SkyBlue}{Integrity Tension (ban vs. use)} & \textcolor{SkyBlue}{Gateway courses} & \textcolor{SkyBlue}{“Can we allow AI and still assess meaningfully and fairly?”} & \textcolor{SkyBlue}{Rubrics, disclosure norms, assessment redesign} \\
\bottomrule
\end{tabularx}

\spacer

\newpage \subsection*{Worksheet 3. Prioritize What to Address First.}

\hdr{Step A: Pick 2--3 high-priority \emph{narratives} from Worksheet 1}\spacer

\noindent \begin{tabularx}{\textwidth}{@{}p{3.8cm} p{2.2cm} Y@{}}
\toprule
\textbf{Narrative} & \textbf{Tier} & \textbf{Why now? (risk/opportunity, urgency)} \\
\midrule
\textcolor{SkyBlue}{AI in the Classroom (assessment redesign)} & \textcolor{SkyBlue}{1} & \textcolor{SkyBlue}{Immediate integrity concerns in Fall; faculty energy is high} \\
\textcolor{SkyBlue}{AI Curriculum (core sequence refresh)} & \textcolor{SkyBlue}{3} & \textcolor{SkyBlue}{Accreditation review next year; employer signals shifting} \\
\bottomrule
\end{tabularx}

\spacer
\hdr{Step B: Pick 2--3 consequential \emph{sentiments} from Worksheet 2}\spacer

\noindent \begin{tabularx}{\textwidth}{@{}p{2.8cm} p{4.4cm} Y@{}}
\toprule
\textbf{Sentiment} & \textbf{Who/where concentrated} & \textbf{Response principle (what they need to hear/see)} \\
\midrule
\textcolor{SkyBlue}{Uncertainty, Confusion} & \textcolor{SkyBlue}{Dept.\ meeting; gateway instructors} & \textcolor{SkyBlue}{Clear staged plan; examples from peers; low-stakes pilots} \\
\textcolor{SkyBlue}{Anxiety, Threat} & \textcolor{SkyBlue}{Methods instructors} & \textcolor{SkyBlue}{Protect core expertise; show what AI can’t or shouldn't replace; scaffold} \\
\bottomrule
\end{tabularx}

\spacer
\hdr{Step C: Map to Tiered Focus}\spacer

\noindent \begin{tabularx}{\textwidth}{@{}p{3.2cm} p{1.2cm} Y Y@{}}
\toprule
\textbf{Priority Area} & \textbf{Tier (1--4)} & \textbf{What success looks like (6--12 months)} & \textbf{First move (next 2-3 months)} \\
\midrule
\textcolor{SkyBlue}{Assessment redesign} & \textcolor{SkyBlue}{1} & \textcolor{SkyBlue}{New AI-aware rubrics; disclosure norms; fewer integrity cases} & \textcolor{SkyBlue}{Workshop on AI literacy + sample rubrics; pilot in two gateway courses} \\
\textcolor{SkyBlue}{Core curriculum refresh} & \textcolor{SkyBlue}{3} & \textcolor{SkyBlue}{AI literacy embedded; signature competencies protected} & \textcolor{SkyBlue}{Map insertion points; propose course-release for leads.} \\
\bottomrule
\end{tabularx}

\spacer

\newpage \subsection*{Worksheet 4. Resources \& Gaps (Be realistic and specific)}
\hdr{Instructions} 
\noindent For each priority, list what you need, what exists, and what’s missing. Add a quick feasibility score (1 = very low, 5 = very high).\spacer

\noindent \begin{tabularx}{\textwidth}{@{}p{2.5cm} Y Y Y p{1.2cm} p{1.6cm}@{}}
\toprule
\textbf{Priority Area} & \textbf{Resources Needed (time, tools, people)} & \textbf{Available Locally (dept)} & \textbf{Available Institutionally} & \textbf{Gap?} & \textbf{Feasib.} \\
\midrule
\textcolor{SkyBlue}{AI pedagogy pilots} & \textcolor{SkyBlue}{Sample assignments; mentor time} & \textcolor{SkyBlue}{2 faculty volunteers} & \textcolor{SkyBlue}{CTL consults; tool licenses} & \textcolor{SkyBlue}{Y} & \textcolor{SkyBlue}{4} \\
\textcolor{SkyBlue}{Core refresh} & \textcolor{SkyBlue}{Course releases; exemplar syllabi} & \textcolor{SkyBlue}{Syllabi archive} & \textcolor{SkyBlue}{seed funds} & \textcolor{SkyBlue}{Y} & \textcolor{SkyBlue}{3} \\
\textcolor{SkyBlue}{Student co-design} & \textcolor{SkyBlue}{Stipends; studio space} & \textcolor{SkyBlue}{Active student club} & \textcolor{SkyBlue}{Innovation hub microgrants} & \textcolor{SkyBlue}{N} & \textcolor{SkyBlue}{5} \\
\bottomrule
\end{tabularx}

\spacer

\subsection*{Worksheet 5. Action Plan by Stage (Flexible, not linear)}
\hdr{Instructions} 

\noindent Use the staged framework as a \emph{map}, not a mandate. Stages may overlap; start where momentum and resources exist.\spacer

\noindent \textbf{Stage 1: AI Literacy (baseline understanding \& critical perspective)}

\noindent \begin{tabularx}{\textwidth}{@{}p{3.4cm} Y p{2.8cm} p{2.2cm} p{1.6cm}@{}}
\toprule
\textbf{Action} & \textbf{Owner(s) \& partners} & \textbf{Start date} & \textbf{Check-in} & \textbf{Status} \\
\midrule
\textcolor{SkyBlue}{Dept-wide primer workshop} & \textcolor{SkyBlue}{CTL + curriculum committee} & \textcolor{SkyBlue}{Oct 1} & \textcolor{SkyBlue}{Nov 15} & \textcolor{SkyBlue}{Planned} \\
\textcolor{SkyBlue}{Reading group (monthly)} & \textcolor{SkyBlue}{Methods faculty + grad reps} & \textcolor{SkyBlue}{Sep} & \textcolor{SkyBlue}{Dec / Mar} & \textcolor{SkyBlue}{On-going} \\
\textcolor{SkyBlue}{Discipline-specific AI brief} & \textcolor{SkyBlue}{Director of Undergrad. Studies} & \textcolor{SkyBlue}{Nov} & \textcolor{SkyBlue}{Jan} & \textcolor{SkyBlue}{Planned} \\
\bottomrule
\end{tabularx}

\spacer
\newpage \noindent \textbf{Stage 2: AI Pedagogy (teaching with/around AI)}

\noindent \begin{tabularx}{\textwidth}{@{}p{3cm} Y p{2.8cm} p{2.2cm} p{1.6cm}@{}}
\toprule
\textbf{Action} & \textbf{Owner(s) \& partners} & \textbf{Start date} & \textbf{Check-in} & \textbf{Status} \\
\midrule
\textcolor{SkyBlue}{Pilot AI Tutor} & \textcolor{SkyBlue}{Course team + CTL} & \textcolor{SkyBlue}{Oct} & \textcolor{SkyBlue}{Dec} & \textcolor{SkyBlue}{Pilot} \\
\textcolor{SkyBlue}{Integrity policy alignment + disclosure} & \textcolor{SkyBlue}{DUS + Education Dean’s office} & \textcolor{SkyBlue}{Oct} & \textcolor{SkyBlue}{Dec} & \textcolor{SkyBlue}{Drafting} \\
\textcolor{SkyBlue}{TA training refresh (AI-aware)} & \textcolor{SkyBlue}{TA coordinator} & \textcolor{SkyBlue}{Nov} & \textcolor{SkyBlue}{Jan} & \textcolor{SkyBlue}{Planned} \\
\bottomrule
\end{tabularx}

\spacer
\noindent \textbf{Stage 3: AI Curriculum (what to teach in an AI-rich era)}

\noindent \begin{tabularx}{\textwidth}{@{}p{3.6cm} Y p{2.8cm} p{2.2cm} p{1.6cm}@{}}
\toprule
\textbf{Action} & \textbf{Owner(s) \& partners} & \textbf{Start date} & \textbf{Check-in} & \textbf{Status} \\
\midrule
\textcolor{SkyBlue}{Map AI literacy across core sequence} & \textcolor{SkyBlue}{Curriculum comm.\ } & \textcolor{SkyBlue}{Oct} & \textcolor{SkyBlue}{Jan} & \textcolor{SkyBlue}{In progress} \\
\textcolor{SkyBlue}{Add AI ethics/limitations module} & \textcolor{SkyBlue}{Methods sequence lead} & \textcolor{SkyBlue}{Jan} & \textcolor{SkyBlue}{Mar} & \textcolor{SkyBlue}{Planned} \\
\textcolor{SkyBlue}{External curr.\ review} & \textcolor{SkyBlue}{Industry advisory board} & \textcolor{SkyBlue}{Feb} & \textcolor{SkyBlue}{Apr} & \textcolor{SkyBlue}{Planned} \\
\bottomrule
\end{tabularx}

\spacer
\noindent \textbf{Stage 4: Research Integration (link teaching $\leftrightarrow$ inquiry)}

\noindent \begin{tabularx}{\textwidth}{@{}p{4.6cm} Y p{2.8cm} p{2.2cm} p{1.6cm}@{}}
\toprule
\textbf{Action} & \textbf{Owner(s) \& partners} & \textbf{Start date} & \textbf{Check-in} & \textbf{Status} \\
\midrule
\textcolor{SkyBlue}{SoTL/disciplinary study on pilots} & \textcolor{SkyBlue}{CTL+ PI} & \textcolor{SkyBlue}{Feb} & \textcolor{SkyBlue}{May} & \textcolor{SkyBlue}{Proposed} \\
\bottomrule
\end{tabularx}

\spacer

\newpage \subsection*{Worksheet 6. Checkpoints \& Governance (Keep it moving, keep it aligned)}

\hdr{Discussion Prompts (to guide reflection and alignment)}  

\hdr{Alignment to Mission of Knowledge (1--2 sentences)}  
\emph{Example: \textcolor{SkyBlue}{“These steps protect our signature competencies while equipping students to act responsibly with AI, sustaining the discipline’s contribution to society.”}}\spacer  

\hdr{Equity \& Integrity Considerations (brief)}  
\emph{Example: \textcolor{SkyBlue}{“Ensure accessible tools; transparent disclosure norms; targeted support for gateway courses.”}}\spacer  

\hdr{Communications Plan (faculty, students, partners)}  
\emph{Example: \textcolor{SkyBlue}{“Monthly update at dept.\ meeting; student town hall; one-pager for advisory board.”}}\spacer  

\noindent \begin{tabularx}{\textwidth}{@{}p{2.8cm} Y Y@{}}
\toprule
\textbf{Checkpoint} & \textbf{Evidence of progress} & \textbf{Decision needed / Escalate to (chair, dean, CTL)} \\
\midrule
\textcolor{SkyBlue}{3 months} & \textcolor{SkyBlue}{Two pilots launched; workshop completed} & \textcolor{SkyBlue}{Approve release time; confirm assessment plan} \\
\textcolor{SkyBlue}{6 months} & \textcolor{SkyBlue}{Interim pilot data; draft curriculum map} & \textcolor{SkyBlue}{Seed funds for next phase; external reviewer list} \\
\textcolor{SkyBlue}{1 year} & \textcolor{SkyBlue}{Revised syllabi; policy adopted;} & \textcolor{SkyBlue}{Scale decisions; institutionalize supports} \\
\bottomrule
\end{tabularx}


\end{document}